\documentclass{article}
\usepackage{spconf,amsmath,graphicx}
\usepackage{array}
\usepackage{multirow}
\usepackage{tabularx}
\usepackage{float}
\usepackage{pdfpages}
\usepackage{amsfonts} 
\usepackage{xcolor}

\title{Phoneme Segmentation Using Self-Supervised Speech Models}
\name{Luke Strgar, David Harwath}
\address{University of Texas at Austin\\Department of Computer Science}
\copyrightnotice{978-1-6654-7189-3/22/\$31.00 ©2023 IEEE}
\begin{document}
%\ninept
%
\maketitle

\begin{abstract}
We apply transfer learning to the task of phoneme segmentation and demonstrate the utility of representations learned in self-supervised pre-training for the task. Our model extends transformer-style encoders with strategically placed convolutions that manipulate features learned in pre-training. Using the TIMIT and Buckeye corpora we train and test the model in the supervised and unsupervised settings. The latter case is accomplished by furnishing a noisy label-set with the predictions of a separate model, it having been trained in an unsupervised fashion. Results indicate our model eclipses previous state-of-the-art performance in both settings and on both datasets. Finally, following observations during published code review and attempts to reproduce past segmentation results, we find a need to disambiguate the definition and implementation of widely-used evaluation metrics. We resolve this ambiguity by delineating two distinct evaluation schemes and describing their nuances. We provide a publicly available implementation of our work on Github \footnote{https://github.com/lstrgar/self-supervised-phone-segmentation}.
\end{abstract}


\section{Introduction}
\label{sec:intro}

Phoneme boundary detection involves labeling the temporal boundaries between discrete phonemic units in a speech signal. Previously, phoneme segmentation has been studied and benchmarked in the supervised \cite{kreuk:segfeat, franke2016phoneme, mcauliffe2017montreal, lin:2022} and unsupervised settings \cite{kreuk:unsupseg, bhati:2021}. In the former case, models are allowed to leverage a ground truth reference segmentation - a vector of phoneme onset, offset times - during training. In the latter case, the model only sees the input speech signal and is thus tasked with producing a segmentation by relying on the statistics of the underlying data alone. A third setting, known as forced-alignment or text-dependent phoneme segmentation, extends the supervised case by adding a temporally ordered list of phonetic identities to the model input. Conditioning on categorical phonetic identity means that model performance in the forced alignment setting typically supersedes text-independent supervised phoneme segmentation, which supersedes unsupervised predictions. In this paper, we focus on and report results for the unsupervised and text-independent supervised cases.

Self-supervised learning is a subclass of unsupervised learning in which training targets are derived from the input data itself. Recently, the speech processing field has benefited from the discovery and refinement of self-supervised strategies. Such heuristic strategies are often employed in a so-called model pre-training phase, and latter pre-trained models are fine-tuned or transfer learning is applied on specific downstream tasks. Numerous speech processing tasks have achieved new state-of-the-art (SotA) performances via application of fine-tuning and transfer learning to the information rich representations learned using self-supervised objectives. These include automatic speech recognition (ASR) \cite{baevski2020wav2vec, hsu2021hubert, yi2020applying}, emotion recognition \cite{mohamed2021arabic, wang2021fine}, and speaker verification \cite{wang2021fine, chen2022large}, among others.

\begin{figure}[t]
\begin{minipage}[b]{0.90\linewidth}
  \centering
  \centerline{\includegraphics[width=0.95\linewidth]{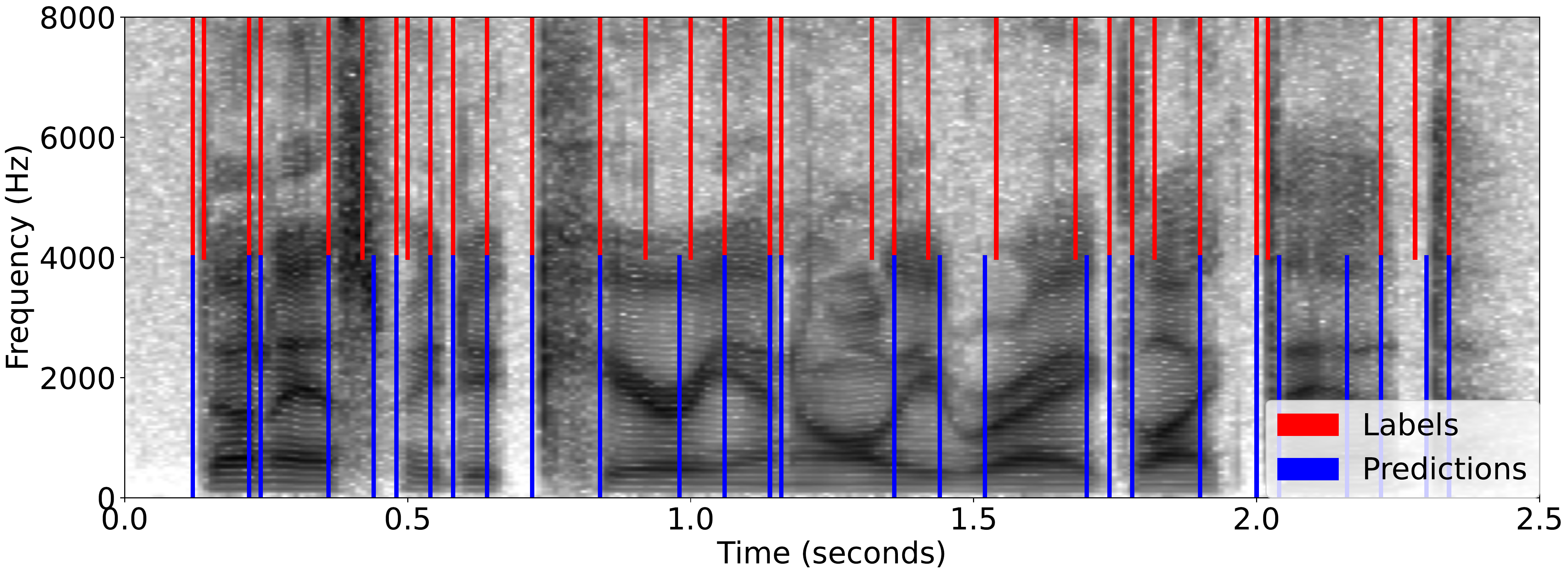}}
  \caption{Example spectrogram with ground truth and supervised model predicted boundaries.}
  \label{fig1}
\end{minipage}
\end{figure}

\begin{figure*}[htb]
\centering
\includegraphics[width=0.95\linewidth]{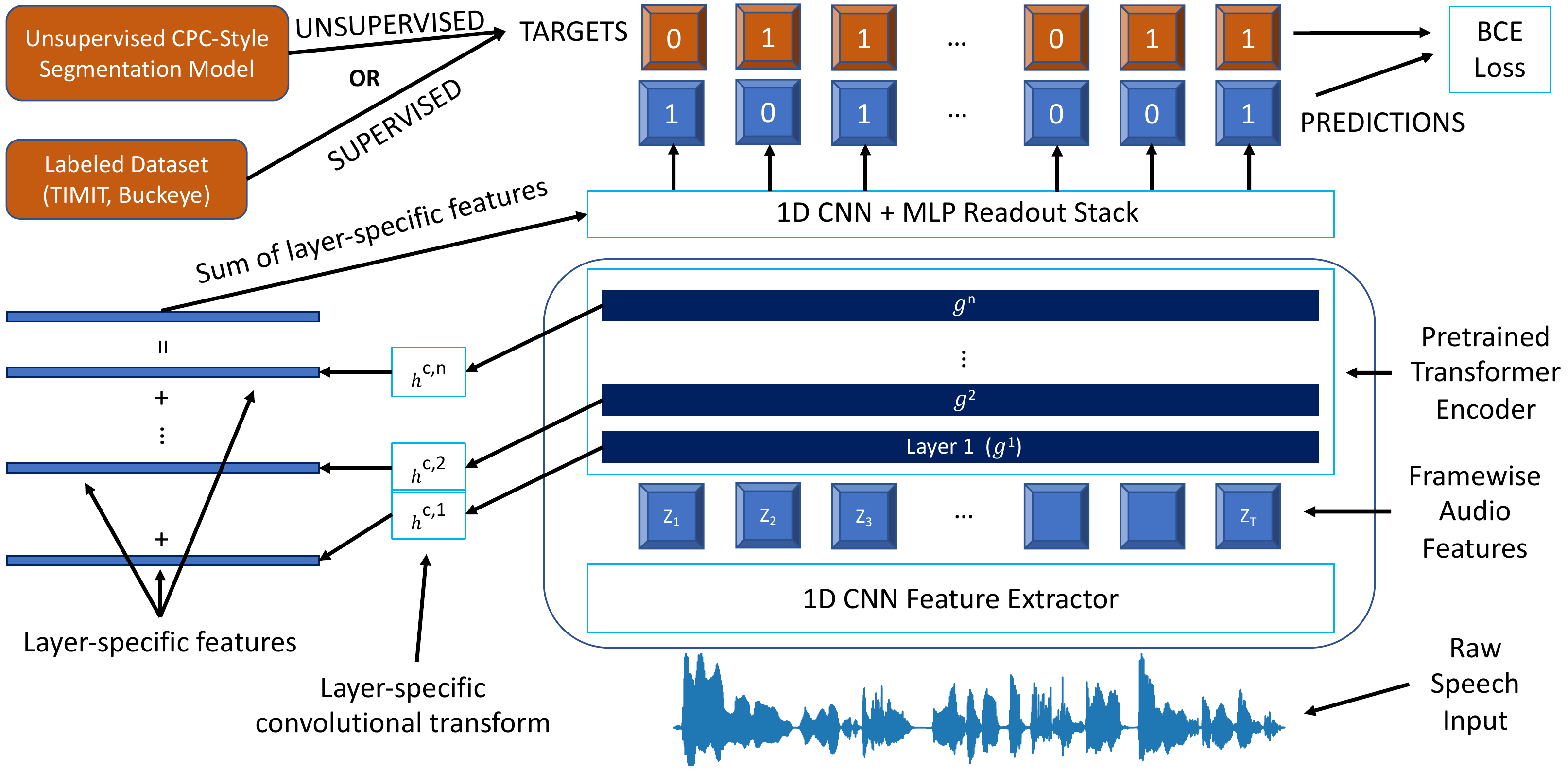}
\caption{\emph{Readout} model architecture schematic. A pre-trained model extracts hierarchical features from the raw waveform. Features are processed by a series of convolutional networks and probability scores are computed. Finally, binary cross entropy loss is evaluated using model predictions and either ground truth labels or noisy labels estimated in an unsupervised manner.}
\label{fig2}
\end{figure*} 

Inspired by the broad successes of self-supervised pre-training in speech processing, in this paper we explore its utility for phoneme segmentation. Specifically, we utilize pre-trained model checkpoints for two well-known and widely used self-supervised speech models, wav2vec2.0 \cite{baevski2020wav2vec} and HuBERT \cite{hsu2021hubert}, and apply different strategies to refine these models' frame-wise representations for phoneme segmentation. In one case, we freeze the model's weights and extend its architecture with strategically placed, trainable, convolutional probe layers to manipulate and synthesize hierarchical features to output a binary predictor for each frame corresponding to the presence of a boundary. In a separate case, we append a simple projection layer to the pre-trained model's encoder and train the projection layer as well as all model weights end-to-end. 

We evaluate and report results using the TIMIT \cite{timit-discacoustic} and Buckeye \cite{pitt2007buckeye} speech corpora and find our model eclipses previous state-of-the-art performance on both datasets in the supervised and unsupervised settings. Unsupervised training is accomplished by furnishing a noisy label-set with the predictions of a separate model \cite{kreuk:unsupseg} that was trained in an unsupervised fashion using contrastive predictive coding \cite{cpc-https://doi.org/10.48550/arxiv.1807.03748} on a next frame prediction task. In supervised training, we also explore the label efficiency of our approach by sweeping over the amount of labeled training data used find that the model surpasses previous SotA performance with as little as 10\% of the training set. With this work we demonstrate a successful application of self-supervised pre-training to the phoneme boundary detection task and offer a new SotA benchmark in the unsupervised and text-independent supervised settings. In addition, in later sections we describe ambiguity and inconsistencies in the commonly used evaluation protocol and offer a resolution in the form of two distinct evaluation schemes.

\begin{table*}[ht!]
    \centering
        \caption{Results obtained in the fully supervised setting. * Indicates application of the strict evaluation framework and ºº denotes author reported scores. The NA placeholder is used where results are not available. Bolded values indicate highest score for the specific metric and dataset.}
    \label{table1}
    \begin{tabular}{|c|c|c|c|c|c|c|c|c|c|}
    \hline
        Data & Model & Precision & Precision* & Recall & Recall* & F1 & F1* & R-Value & R-Value* \\\cline{1-10}
        \multirow{7}{*}{Buckeye} & Lin et al. \cite{lin:2022} ºº & 88.49 & NA & 90.33 & NA & 89.40 & NA & 90.90 & NA \\\cline{2-10}
        & Kreuk et al. \cite{kreuk:segfeat} ºº & 85.40 & NA & 89.12 & NA & 87.23 & NA & 88.76 & NA\\\cline{2-10}
        & W2V2 finetune & \textbf{94.01} & \textbf{90.56} & 93.08 & \textbf{90.28} & \textbf{93.54} & \textbf{90.42} & \textbf{94.41} & \textbf{91.81} \\\cline{2-10}
        & HuBERT finetune & 93.83 & 89.81 & \textbf{93.11} & \textbf{90.28} & 93.47 & 90.05 & 94.37 & 91.51 \\\cline{2-10}
        & W2V2 readout & 93.38 & 89.14 & 92.74 & 89.66 & 93.00 & 89.40 & 93.99 & 90.96 \\\cline{2-10}
        & HuBERT readout & 93.37 & 89.30 & 92.95 & 89.94 & 93.16 & 89.62 & 94.13 & 91.15 \\\cline{1-10}
        \hline
        \hline
        \multirow{7}{*}{TIMIT} & Lin et al. \cite{lin:2022} ºº & 93.42 & NA & 95.96 & NA & 94.67 & NA & 95.18 & NA \\\cline{2-10}
        & Kreuk et al. \cite{kreuk:segfeat} ºº & 94.03 & NA & 90.46 & NA & 92.22 & NA & 92.79 & NA \\\cline{2-10}
        & Kreuk et al. \cite{kreuk:segfeat} & 92.94 & 92.14 & 92.31 & 89.26 & 92.63 & 90.68 & 93.66 & 91.71 \\\cline{2-10}
        & W2V2 finetune & 96.90 & \textbf{94.35} & \textbf{96.30} & \textbf{93.91} & \textbf{96.60} & \textbf{94.13} & \textbf{97.04} & \textbf{94.96} \\\cline{2-10}
        & HuBERT finetune & \textbf{96.93} & 94.31 & 96.09 & 93.68 & 96.51 & 94.00 & 96.92 & 94.83 \\\cline{2-10}
        & W2V2 readout & 96.67 & 93.75 & 95.56 & 92.65 & 96.11 & 93.20 & 96.55 & 94.10 \\\cline{2-10}
        & HuBERT readout & 96.50 & 93.23 & 95.93 & 93.47 & 96.21 & 93.35 & 96.71 & 94.33 \\\cline{1-10}
    \end{tabular}
\end{table*}

\section{Related Work}
\label{sec:related-work}

\subsection{Phoneme Boundary Detection}
Phoneme boundary detection has been explored using a variety of different model types and under various levels of supervision. In the text-independent, supervised setting, recent work revolves around the usage of recurrent neural network models. RNNs have been used as binary predictors \cite{franke2016phoneme} and feature learners for a subsequent structured prediction task \cite{kreuk:segfeat}. In text-dependent phoneme segmentation, probabilistic models such as HMMs have been applied \cite{mcauliffe2017montreal}, and recently a multi-task learning framework using pre-trained model features was proposed \cite{lin:2022}. In the unsupervised setting, signal processing based approaches were initially dominant~\cite{glass03, dusan06}, but recent research has focused on learning-based methods. \cite{lee-glass-2012-nonparametric} proposed a nonparametric Bayesian approach to unsupervised phonetic segmentation and clustering, and more recently the noise contrastive estimation principle has been applied to optimize the similarity of adjacent frames while making distant frames dissimilar \cite{kreuk:unsupseg}. Other work has applied contrastive learning at multiple levels by jointly optimizing both phoneme and word segmentation models \cite{bhati:2021}. 

\subsection{Self-Supervised Pre-Training}
Self-supervised pre-training has seen great success in numerous speech processing tasks. Borrowing ideas from research in natural language processing and computer vision, self-supervised models such as wav2vec2.0 \cite{baevski2020wav2vec} and HuBERT \cite{hsu2021hubert} are trained to reconstruct masked input from unmasked representations. The resulting internal representations obtained by these and other training objectives have been successfully applied to downstream tasks including ASR \cite{baevski2020wav2vec, hsu2021hubert, yi2020applying}, emotion recognition \cite{mohamed2021arabic, wang2021fine}, and speaker verification \cite{wang2021fine, chen2022large}, among others.

\section{Problem Statement}
\label{sec:problem-and-eval}

In phoneme segmentation the input is a raw speech waveform $x \in \mathcal{X}$ represented as $x = (x_0, x_1, ..., x_N)$ where each $x_i$ is a single floating point value representing relative pressure in the transmission medium. Typically, $x$ will be pre-processed and temporally down-sampled by some transformation $f_x: \mathcal{X} \rightarrow \mathcal{Z}$ to produce $f_{x}(x) = z = (z_1, z_2, ..., z_T)$ where $T << N$ and $z_i \in \mathbb{R}^d$. Here, $z \in \mathbb{R}^{T x d}$ can be thought of as representing a series of acoustic feature frames and $T$ now encodes the temporal resolution we desire to make predictions with. 

Each input speech sample is paired with a label sequence of time stamps $y = (y_1, y_2, ..., y_K)$ where each $y_i$ indicates the presence of one boundary and is represented as a single floating point value encoding the time units relative to the beginning of the utterance. Similar to the down-sampling of $x$, one might choose to bin $y$ such that each element is converted to units of acoustic feature frames. We call this representation $\bar{y} = (\bar{y}_1, ..., \bar{y}_K)$ and note that $K$, $N$, and $T$ may vary across input / label pairs. 

Automatic phoneme segmentation thus asks for a prediction, $\hat{y} = (\hat{y}_1, \hat{y}_2, ..., \hat{y}_{\hat{K}})$ that closely matches the ground truth label $\bar{y}$. Classically, the closeness of a reference and predicted segmentation is evaluated with the precision, recall, F1, and R-Value metrics \cite{rasanen2009improved}. Section 5 describes these quantities as well as their nuances in detail. 

\section{Model Description}
\label{sec:model-description}

Inspired by the success of self-supervised learning in numerous speech processing tasks we adopt pre-trained wav2vec2.0 and HuBERT model checkpoints to compose the backbone of a frame-wise binary classifier. Both pre-trained models share a similar architecture. For our purposes we consider only the encoder, which we denote by the function composition $g \circ f$. Elaborating on the component functions of this composition, $f: \mathcal{X} \rightarrow \mathcal{Z}$ is commonly referred to as the convolutional feature extractor, which processes raw waveform input and outputs a time series of latent speech representations. Thus, $f$ acts like the previously defined $f_x$; however, $f$ is not strictly a pre-processing step since it is learned during end-to-end training of wav2vec2.0 and HuBERT. Meanwhile, $g: \mathcal{Z} \rightarrow \mathcal{C}$ is known as the context network, which applies learned attention masks to synthesize a context-aware representation $c_i \in \mathcal{C}$ from each $z_i \in \mathcal{Z}$. $g$ is itself a compositional function built from a cascade of $n$ transformer self-attention blocks. Thus, we can also write $g = g^{n} \circ g^{n-1} \circ... \circ  g^1$. Note that functions $f$ and $g$ may be initialized by either wav2vec2.0 or HuBERT.

\begin{figure}[htb]
\centering
\includegraphics[width=0.95\linewidth]{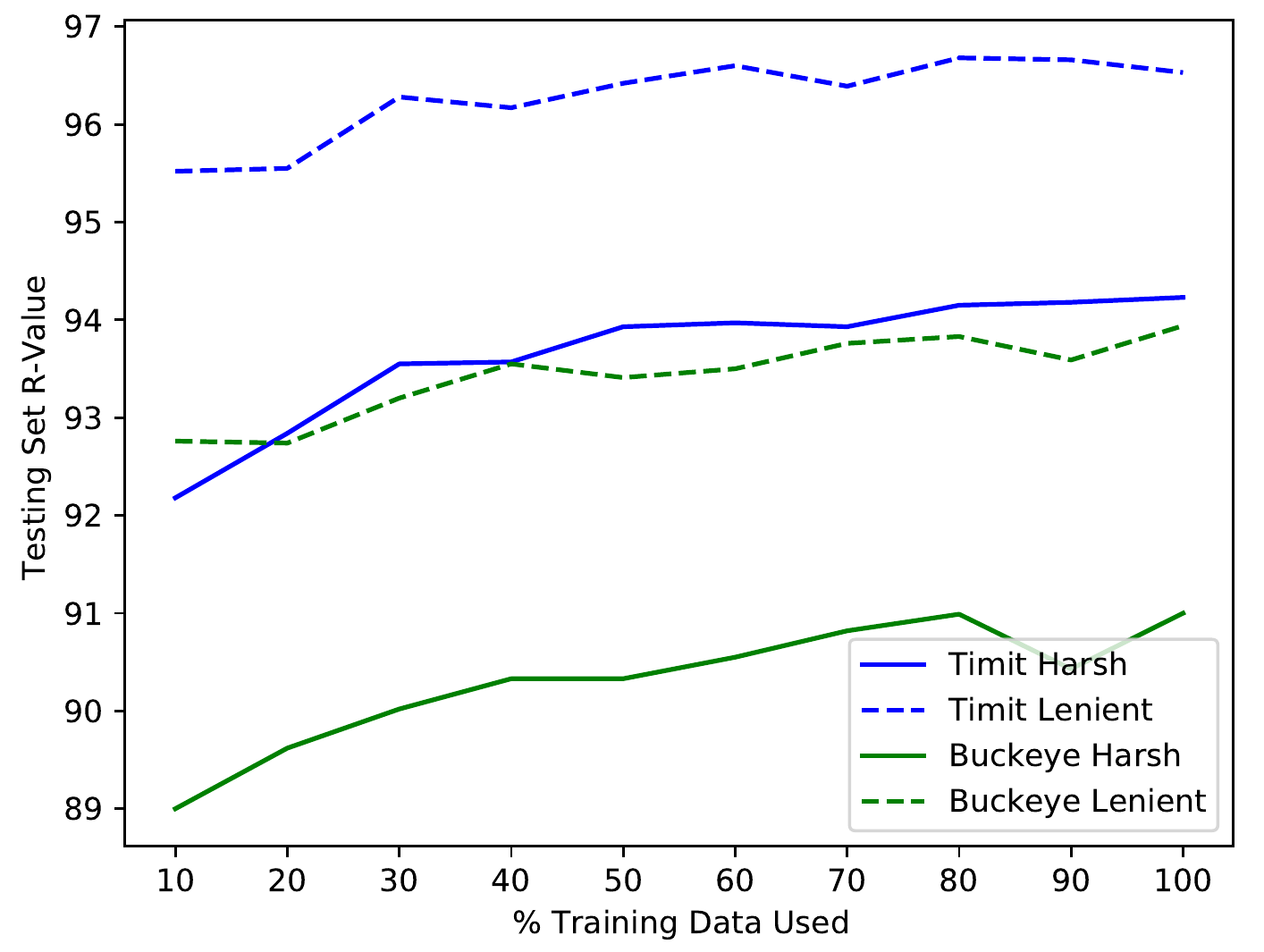}
\caption{Supervised model with wav2vec2.0 backbone in \emph{readout} mode trained from scratch on incrementally larger fractions of labeled data. Vertical axis shows testing set R-Value performance.}
\label{fig3}
\end{figure} 

We develop two separate classification model formulations built on-top of the pre-trained network backbone. The first case, which we call \emph{fine-tune} mode, appends a single linear projection layer, $h^{ft}$, to the output of the pre-trained model. As the name suggests, in this setting, the entire pre-trained model and added projection receive gradient updates, and the model can be formalized as the function composition $f \circ g \circ h^{ft}$.

The second case, called \emph{readout} mode, is depicted in Figure~\ref{fig2}. Here, we freeze the pre-trained model and apply learned, layer-specific convolutions, $h^{c,1}, h^{c,2}, ..., h^{c,n}$, to feature representations extracted from each $g^{i}$. The outputs are then summed and passed through a final series of convolutions and perceptron layers, denoted $h^{ro}$. Empirically, we discovered that applying a learned weight parameter to each layer's processed features before computing the summation improved performance; however, we omit these terms in the following expression for simplicity. Denoting the outputs of each $g^{i}$ as $c^{i}$, the \emph{readout} model can be formalized as $h^{ro}(\sum_{i} h^{c,i}(c^i))$. .

 Both models output a series of frame-wise binary labels, $\hat{y}_b$, where a $1$ is interpreted as the occurrence of a boundary. Given a training set of input utterances $\mathcal{S} = \ \{x^i, y_{b}^{i}\}_{i=1}^{m}$, loss is computed and models are updated according to the binary cross entropy (BCE) objective function in Equation~\ref{equation1}. Here, the term $w^*$ is a strictly positive weight value assigned to the loss associated with frames where the reference ground truth indicates the presence of a boundary.  
 
 % \vspace{-1cm}
\begin{equation}
    \mathcal{L}_{BCE} = \sum_{i=0}^{m} \sum_{j \in y_{b}^{i}} w^* y_{b, j}^{i} \log(\hat{y}_{b,j}) + (1 - y_{b, j}^{i}) \log(1 - \hat{y}_{b,j})
    \label{equation1}
\end{equation}

Optimization of the above objective is explored in the supervised and unsupervised settings. In the former case, we rely on the time-aligned transcriptions that come as part of the TIMIT and Buckeye corpora to construct supervised targets. In the latter case, we perform inference over the TIMIT and Buckeye training sets using the unsupervised model provided by Kreuk et al. \cite{kreuk:unsupseg}. These predicted labels then serve as the supervisory signal used during training with BCE loss.

\begin{table*}[ht!]
    \centering
        \caption{Results obtained in the unsupervised setting. A noisy label-set was furnished using publicly available checkpoints from an unsupervised segmentation model \cite{kreuk:unsupseg}.}
    \label{table2}
    \begin{tabular}{|c|c|c|c|c|c|c|c|c|c|}
    \hline
        Data & Model & Precision & Precision* & Recall & Recall* & F1 & F1* & R-Value & R-Value* \\\cline{1-10}
        \multirow{7}{*}{Buckeye} & Bhati et al. \cite{bhati:2021} ºº & 76.53 & NA & 78.72 & NA & 77.61 & NA & 80.72 & NA \\\cline{2-10}
        & Kreuk et al. \cite{kreuk:unsupseg} ºº & 75.78 & NA & 76.86 & NA & 76.31 & NA & 79.69 & NA \\\cline{2-10}
        & Kreuk et al. \cite{kreuk:unsupseg} & 77.17 & 72.21 & 79.71 & 75.55 & 78.42 & 73.85 & 81.39 & 77.28 \\\cline{2-10}
        & W2V2 finetune & 82.15 & 75.56 & \textbf{85.13} & \textbf{79.47} & 83.61 & 77.47 & 85.81 & 80.33 \\\cline{2-10}
        & HuBERT finetune & 83.09 & 76.62 & 84.47 & 78.75 & 83.77 & \textbf{77.67} & 86.11 & 80.79 \\\cline{2-10}
        & W2V2 readout & \textbf{84.24} & \textbf{77.92} & 82.88 & 77.41 & 83.55 & \textbf{77.67} & 85.92 & \textbf{80.95} \\\cline{2-10}
        & HuBERT readout & 83.35 & 75.29 & 84.68 & 79.37 & \textbf{84.01} & 77.28 & \textbf{86.31} & 80.13 \\\cline{1-10}
        \hline
        \hline
        \multirow{7}{*}{TIMIT} & Bhati et al. \cite{bhati:2021} ºº & 84.63 & NA & 86.04 & NA & 85.33 & NA & 87.44 & NA \\\cline{2-10}
        & Kreuk et al. \cite{kreuk:unsupseg} ºº & 83.89 & NA & 83.55 & NA & 83.71 & NA & 86.02 & NA \\\cline{2-10}
        & Kreuk et al. \cite{kreuk:unsupseg} & 85.27 & 81.42 & 83.48 & 76.53 & 84.36 & 78.90 & 86.57 & 81.71 \\\cline{2-10}
        & W2V2 finetune & 88.93 & 82.16 & \textbf{88.60} & 80.83 & 88.76 & 81.49 & 90.40 & 84.18 \\\cline{2-10}
        & HuBERT finetune & 89.05 & 82.07 & 88.44 & 80.70 & 88.75 & 81.38 & 90.37 & 84.08 \\\cline{2-10}
        & W2V2 readout & 90.69 & \textbf{84.92} & 86.78 & 78.52 & 88.69 & 81.59 & 89.90 & 83.69 \\\cline{2-10}
        & HuBERT readout & \textbf{90.98} & 82.44 & 88.48 & \textbf{81.18} & \textbf{89.71} & \textbf{81.81} & \textbf{90.98} & \textbf{84.45} \\\cline{1-10}
    \end{tabular}
\end{table*}

\section{Evaluation}
\label{sec:evaluation}

Previous work has articulated the challenges of conducting representative evaluations of phoneme segmentation results \cite{rasanen2009improved}. These issues are further confounded by the wide range of (prediction to label) temporal tolerance levels found in the literature for true positives identification. Here, we consider a 20 millisecond tolerance window on either side of a ground truth label for true positive calling, which is consistent with recent work \cite{kreuk:segfeat, kreuk:unsupseg, bhati:2021, lin:2022}. 

We report our results in terms of precision, recall, F1, and R-Value according to their definition in \cite{rasanen2009improved}. These metrics and their interpretation are widely cited in the phoneme segmentation task literature; however, based on a combination of published code review and attempts to reproduce results we believe there remains meaningful ambiguity in the calculation of precision, recall and their derivative quantities (e.g., F1 and R-Value). By elaborating on these definitions and their implications below, we hope to align the community around shared standards for reporting phoneme boundary detection results. 

\begin{figure}[htb]
\centering
\includegraphics[width=0.95\linewidth]{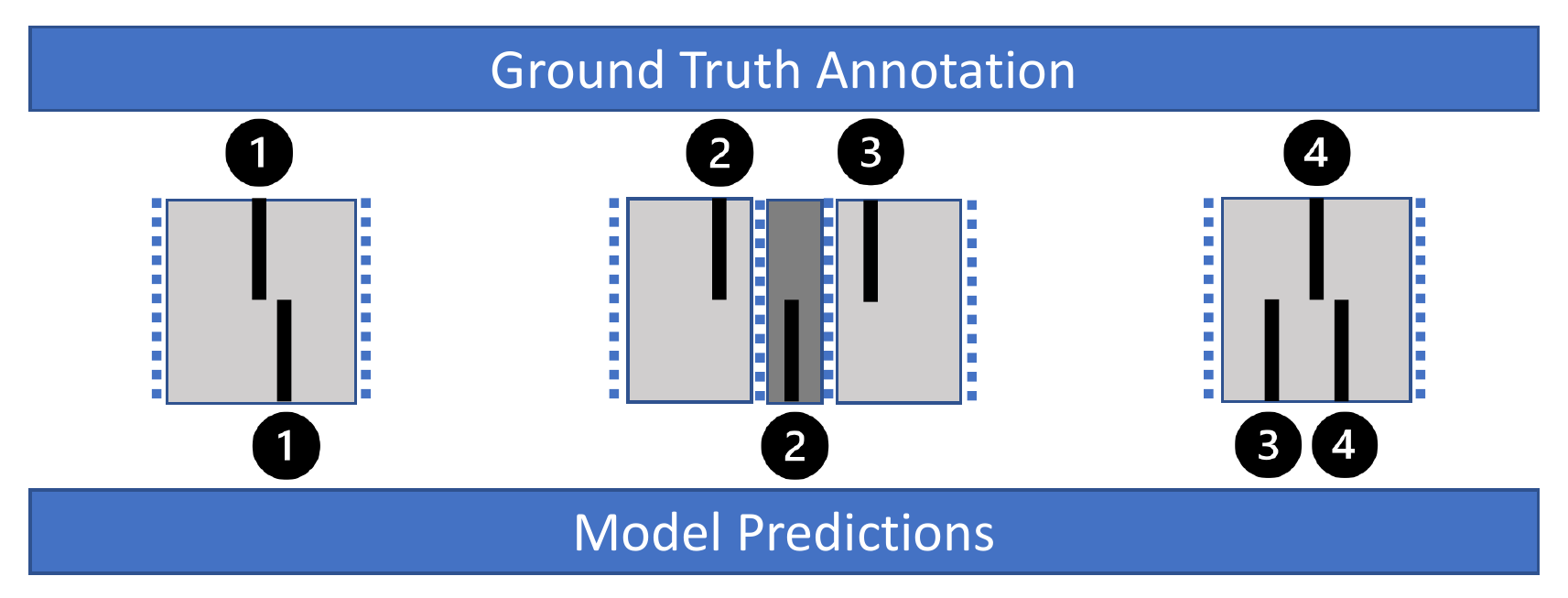}
\caption{Illustration of ambiguities during phoneme segmentation evaluation. Vertical black stripes indicate ground truth (top) and predicted (bottom) boundaries. The light gray regions correspond to ground truth boundary tolerance windows and the dark gray region shows where two tolerance windows overlap. Predicted and ground truth boundaries 1 match. Ground truth boundaries 2, 3 both match predicted boundary 2 while predicted boundaries 3 and 4 both match and ground truth 4.}
\label{fig4}
\end{figure}

When computing precision the primary source of ambiguity revolves around interpreting multiple positive boundary predictions falling within the tolerance window of a ground truth boundary. For recall, the parallel situation arises where a single predicted positive boundary falls within the tolerance window of more than one ground truth boundary. See Figure~\ref{fig4} for a visual illustration of the ambiguous situations arising during evaluation. 

Without loss of generality, consider a boundary predictor operating at 50Hz (i.e. predictions correspond to the occurrence of a boundary in a 20 millisecond window). While computing the precision of this model's predictions with a 20 millisecond tolerance window, one could encounter an isolated ground truth boundary at frame $p$ and model boundary predictions at $p-1$, $p$, $p+1$. Different approaches to this calculation could result in a three-fold difference in performance, and the situation would be exacerbated by an increase in the predictor's frame rate. In practice, the statistics of English language phoneme duration and presentation render a three-fold performance difference highly unlikely; however, others have reported differences of up to 5\% \cite{rasanen2007speech}, and our results consistently show deviations of 3-4\% in the supervised setting and 5-7\% in the unsupervised setting. 

We then wish to delineate a \emph{strict} and \emph{lenient} evaluation scheme for phoneme boundary detection where the \emph{strict} scheme prohibits double counting and the \emph{lenient} scheme allows it. Specifically, while computing the hit rate \cite{rasanen2009improved} of an automated phoneme segmenter, in the \emph{strict} scheme once a ground truth boundary is matched by a predicted boundary the ground truth is removed from consideration for matching additional model predictions. On the other hand, in the \emph{lenient} scheme the same ground truth boundary may match multiple predicted boundaries so long as they fall within the tolerance window. Further, in the \emph{lenient} scheme, the hit rate used for computing precision and that for recall may differ since more than one predicted boundary is allowed to match a the same ground truth and visa versa. 

We denote results following the \emph{strict} scheme with a * and then define F1* and R-Value* as those metrics computed with their \emph{strict} counterparts P* (precision*) and R* (recall*). Our code reviews and efforts to reproduce past results indicate that previous SotA methods use the \emph{lenient} scheme. For parity, our results tables below include both \emph{strict} and \emph{lenient} scores for our models. In some cases where we were able to reproduce previous published results we also add \emph{strict} scores. In other cases, it was not possible to verify the exact evaluation framework used by some authors. However, all these papers explicitly describe sharing evaluation methodology with the aforementioned previous SotA, against which they benchmark their model performance. Accordingly, we assume they also evaluate performance using the \emph{lenient} framework. 

\section{Experiments}
\label{sec:experiments}

\subsection{Datasets}
\label{ssec:datasets} We used the TIMIT \cite{timit-discacoustic} and Buckeye \cite{pitt2007buckeye} speech corpora to train and evaluate the \emph{fine-tune} and \emph{readout} models. For TIMIT, we used the standard train/test split and sampled 10\% of the training data for model validation. For Buckeye, we followed previous work \cite{kreuk:segfeat, kreuk:unsupseg, franke2016phoneme} in our training, validation, and testing set construction. First, we split the corpus at the speaker level, reserving 80\%, 10\%, 10\% for training, validation, and testing, respectively. In addition, long recordings were split during non-vocal noise and silence into shorter continuous speech segments such that each segment starts and ends with no more than 20 milliseconds of non-speech. 

\subsection{Experimental Setup}
\label{ssec:experimental-setup}

Experiments conducted with HuBERT used the base architecture and those with wav2vec2.0 used the small architecture. Both model checkpoints were pre-trained on Librispeech~\cite{Panayotov2015LibrispeechAA} and collected from Fairseq \cite{ott2019fairseq}. We explored the effectiveness of larger model architectures (e.g. wav2vec2.0 large, HuBERT large/x-large) but found they offered no boost on final performance metrics. For our unsupervised experiments, we used model checkpoints made available with the code accompanying \cite{kreuk:unsupseg} to bootstrap labels for TIMIT and Buckeye. 

All models were trained on an NVIDIA Quadro RTX 8000 with a batch size of 16 for 50 epochs. The Adam optimizer was used with a learning rate of 1e-3 and 1e-4 while training in \emph{readout} and \emph{fine-tune} mode, respectively. Models were regularly evaluated during training using the validation set's R-Value* and the best performing model was saved for testing. 

In \emph{readout} mode the layer specific convolutions, $h^{c,i}$ were defined with a kernel size of 9, stride of 1, and 768 input and channels. The output architecture $h^{ro}$ is a depth five convolution stack with a shared kernel size of 3 and stride of 1 followed by a linear projection. As we mentioned previously, in this setting we also added a parameter to learn a weighted sum of the layer specific features before application of $h^{ro}$. 

Throughout our experiments we explored various values of $w^*$ - the loss weight applied to frames labeled as boundary positive. In all supervised experiments, $w^*$ was ultimately set to 1 for the entire duration of model training. We made anecdotal observations that setting $1 < w^* < 2.5$ tended to speed up model convergence; however, $w^*$ had to be subsequently turned down and training continued to obtain the best performance metrics. In the unsupervised setting, we found that, relative to ground truth labels, the noisy labels scored substantially lower in recall than precision. Acknowledging the need then to incentive positive predictions, we swept values of $w^*$ and obtained optimal validation performance using $w^* = 1.4$.

\subsection{Results}

\label{ssec:results}
In Table~\ref{table1} we report results for our models in the fully supervised setting. We also include reported scores from Lin et al. \cite{lin:2022}, Kreuk et al. \cite{kreuk:segfeat}, which stand as previous benchmark results in text-dependent and text-independent phoneme segmentation, respectively. Another result we include is our attempt at reproducing Kreuk et al.'s \cite{kreuk:segfeat} results for TIMIT - here we are able to share both the \emph{harsh} and \emph{lenient} evaluations. We were unable to reproduce comparable scores for Buckeye using the model from \cite{kreuk:segfeat}. Altogether, results indicate that the best of our four models - composed through a selection of a backbone pre-trained network and \emph{fine-tune} or \emph{readout} mode - eclipse previous SotA in every metric category for both TIMIT and Buckeye. With few exceptions, all four of our models outpace previous SotA, and we emphasize that our top performing model, which was trained in the text-independent setting, surpasses the performance of SotA text-dependent \cite{lin:2022}.

Figure~\ref{fig3} highlights the small amount of labeled training data required to surpass previous SotA performance. Results reported in this figure come from experiments with a \emph{readout} mode model trained with a wav2vec2.0 back-bone. For both TIMIT and Buckeye we obtain R-Value SotA using only 10\% of the labeled data from the respective training sets.  

Table~\ref{table2} reports results for models in the unsupervised setting along with other previous SotA results. As in the supervised case, our best performing model achieves a new SotA result for both TIMIT and Buckeye in every metric category. Notably, wherein the supervised setting a typical deviation between the \emph{lenient} and \emph{harsh} schemes is in the 2-3\% range, in the unsupervised setting we observe deviations of, in some cases, more than 8\%. As the Kreuk et al. \cite{kreuk:unsupseg} and Bhati et al. \cite{bhati:2021} unsupervised models reported here perform inference through a peak-picking algorithm over a learned representation, it is possible that over prediction near boundaries stems from the difficulty of enforcing temporally precise transitions in the learned representation. Similarly, as our models are trained using a noisy label-set boostrapped from \cite{kreuk:unsupseg}, our model is liable to the same failure modes.

During experiments with noisy (unsupervised) label-sets, we explored the impact of multiple self-training loops to refine the labels and improve final model test performance. Ultimately, we observed marginal gains that did not inspire a deep exploration of how bootstrapped labels could be refined in an unsupervised fashion. In fact, in \emph{fine-tune} mode, performance declined after multiple self-training loops. Incidentally, throughout our experiments in the unsupervised setting, \emph{readout} models tended to perform better than their fully fine-tuned counterparts. Relevant metrics observed during training indicated that the more expressive fine-tuned models were much more liable to over-fit label noise than their \emph{readout} mode counterparts.

\section{Discussion}
\label{sec:discussion}

% \begin{itemize}
%     \item Recap of contribution and results
%     \item Directions for future work: self-supervised objectives, multi-lingual setting
%     \item 
% \end{itemize}

Here we introduced a new model formulation based on self-supervised pre-training and transfer learning to perform phoneme boundary detection in the supervised and unsupervised settings. We empirically demonstrate that our formulation sets a new SotA benchmark for both settings on standard datasets used for the task - the TIMIT and Buckeye speech corpora. Additionally, we bring to the community's attention a need for shared implementation strategies for key evaluation metrics and define two evaluation frameworks that can be used to alleviate future ambiguity. 

We believe there are several promising directions for future work. First, an exploration of regularization and self-training strategies to improve noisy label-sets will likely push unsupervised results further than we have been able to. Second, in the supervised setting we obtained excellent performance even with small amounts of training data. We are optimistic then that low resource languages can benefit from self-supervised pre-training for phoneme boundary detection. Finally, our model formulation may be, with minimal modifications, well-suited to alternate speech segmentation tasks.

% Below is an example of how to insert images. Delete the ``\vspace'' line,
% uncomment the preceding line ``\centerline...'' and replace ``imageX.ps''
% with a suitable PostScript file name.
% -------------------------------------------------------------------------
% \begin{figure}[htb]

% \begin{minipage}[b]{1.0\linewidth}
%   \centering
%   \centerline{\includegraphics[width=8.5cm]{image1}}
% %  \vspace{2.0cm}
%   \centerline{(a) Result 1}\medskip
% \end{minipage}
% %
% \begin{minipage}[b]{.48\linewidth}
%   \centering
%   \centerline{\includegraphics[width=4.0cm]{image3}}
% %  \vspace{1.5cm}
%   \centerline{(b) Results 3}\medskip
% \end{minipage}
% \hfill
% \begin{minipage}[b]{0.48\linewidth}
%   \centering
%   \centerline{\includegraphics[width=4.0cm]{image4}}
% %  \vspace{1.5cm}
%   \centerline{(c) Result 4}\medskip
% \end{minipage}
% %
% \caption{Example of placing a figure with experimental results.}
% \label{fig:res}
% %
% \end{figure}

%\vfill
%\pagebreak

\bibliographystyle{IEEEbib}
\bibliography{refs}

\end{document}